\documentclass[a4paper]{spie}  
\usepackage[dvips]{graphicx}
\usepackage{epsfig}
\usepackage{amssymb, amsmath, subfigure}

\title{Stochastic evolution and multifractal classification of prokaryotes} 

\author{Matthew J. Berryman\supit{a}, Andrew Allison\supit{a}, and Derek Abbott\supit{a}
\skiplinehalf
\supit{a}Center for Biomedical Engineering and\\
School of Electrical and Electronic Engineering,\\
The University of Adelaide, SA  5005, Australia}

\authorinfo{Send correspondence to Derek Abbott\\
E-mail: dabbott@eleceng.adelaide.edu.au, Telephone: +61 8 8303 5748}

\begin{document} 
\maketitle
\begin{abstract}
We introduce a model for simulating mutation of prokaryote DNA sequences. Using that model we can then evaluated traditional techniques like parsimony and maximum likelihood methods for computing phylogenetic relationships. We also use the model to mimic large scale genomic changes, and use this to evaluate multifractal and related information theory techniques which take into account these large changes in determining phylogenetic relationships.
\end{abstract}

\section{INTRODUCTION}
In this paper we examine the stochastic evolution of prokaryotes by simulating a number of mutational events from changes within genes to changes in overall genome structure. We then use the resultant {\it in silico} virtual mutants~\cite{mut1}, for which we know the ancestry, to compare the accuracy of both whole genome comparisons and more traditional orthologous gene comparisons in deriving phylogenetic relationships.

The relationships between various prokaryotes, bacteria and archaea, are of great interest~\cite{history,pathogens}. 
Since the seminal work by Zuckerkandl and Pauling~\cite{ZuckerkandlPauling}, many of
these relationships have been explored by comparing the DNA and amino acid sequences of the organisms in question. Given a set of sequences from several species, we would then like to infer (with
statistical significance) a phylogenetic relationship between the species. A critical assumption is
that the sequences diverged from a common ancestor and not by other processes such as gene duplication~\cite{GeneDuplication} or gene transfer~\cite{genetransfer}, these are known as orthologues.
The algorithm used to analyze the sequence needs to produce a measure of divergence from the common ancestor which can then be used to construct a variety of trees~\cite{BSA}.

A number of processes are well known by which prokaryotes can mutate and thus diverge from a common ancestor~\cite{ProkaryoticGenetics}. The main mechanisms we have focused on are:
\begin{itemize}
\item Base substitutions, where one base pair has been replaced with a different base through some mechanism (such as UV irradiation with an absent or unsuccessful repair process).
\item Additions and deletions, where a base pair has been added or removed from the sequence.
\item Rearrangements, where a sequence has been inverted or shifted to another location (or both).
\item Gene transfer, where one or more genes are inserted into a prokaryote genome from another source (such as a phage or by recombination with a plasmid).
\end{itemize}

In addition to analyzing orthologous genes, one can compare whole genomes and their features and properties due to the large number of complete prokaryote genomes available~\cite{multiplying}. There have been a number of methods proposed for using a whole genome approach to problems of phylogeny. These include measuring the fraction of orthologs shared between genomes and quantifying correlations between genes with respect to their relative positions in genomes~\cite{GenomeEvolution}. Similarly, comparing orderings of genes between genomes has been proposed as an area of investigation~\cite{GeneOrder}. Other techniques which we have explored in this paper include a multifractal approach~\cite{Multifractal2} and related information theory approaches~\cite{Gutman,InfoDist}.
\section{RESULTS}
\subsection{Comparison of standard algorithms}
We have compared a known tree that we have generated by simulating mutation of the human adenosine a2 receptor~\cite{a2a_receptor} with trees produced by standard techniques such as maximum likelihood~\cite{ml,BSA} and maximum parsimony~\cite{BSA}.

The basic algorithm of our mutation simulation software is as follows: 
\begin{enumerate}
\item Take the original nucleotide sequence, and apply up to five insertions, up to five deletions, and up to five base substitutions. We use a small random number of each to simulate random mutations and to put some
distance between each of the generations, but not too much that it becomes too easy for the algorithms to 
distinguish between them.
Repeat this step $n$ times to produce $n$ descendants of the original, these we called $0,\ldots,n-1$.
\item Taking the mutations of the previous step, ${x}$, we then mutate these to produce some $m$ descendants of those descendants, which we label $x\oplus0, \ldots, x\oplus(m-1) \forall{x}$ (where $\oplus$ denotes concatenation).
\item Repeat the previous step several times.
\end{enumerate}

We then use the CLUSTALW software~\cite{CLUSTALW1,CLUSTALW2} to perform the alignments. The algorithm, as described in Durbin {\it et al.}~\cite{BSA} is:
\begin{enumerate}
\item Construct a distance matrix of all pairs of sequences by a pairwise dynamic programming alignment, followed by conversion of similarity scores to approximate evolutionary distances using the Kimura model~\cite{Kimura}.
\item Construct a guide tree by a neighbor-joining clustering algorithm of Saitou and Nei~\cite{SaitouNei}.
\item Progressively align at nodes in order of decreasing similarity, using sequence-sequence, sequence-profile, and profile-profile alignment.
\end{enumerate}

\begin{figure}[htbp]
  \centering
  \subfigure[Tree generated using the parsimony method.]{
  \epsfig{file=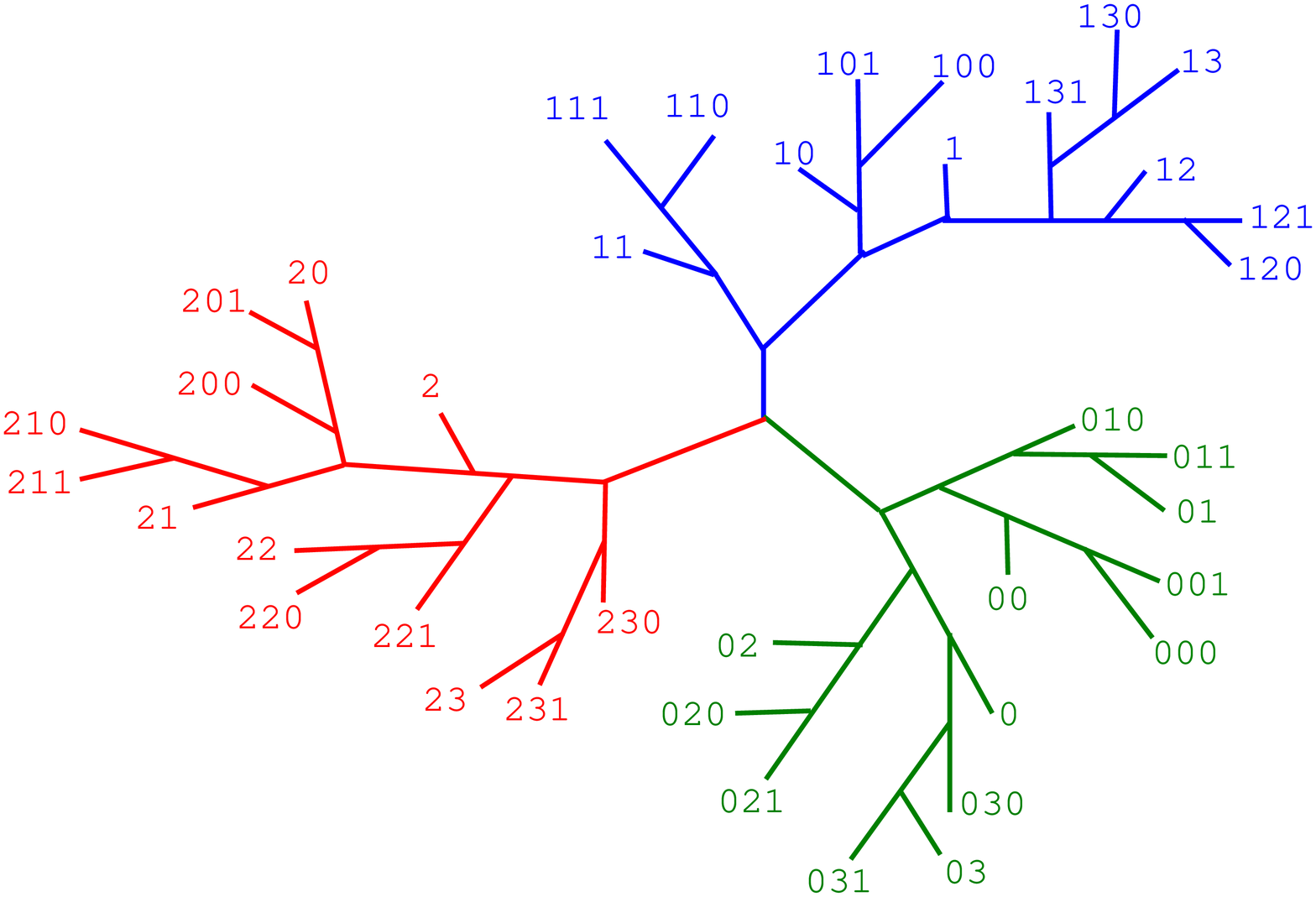,height=7cm}
  \label{trees:parsimony}}
  \subfigure[Tree generated using the maximum likelihood method. The tree is constructed to give a maximum likelihood score, and for this tree the log likelihood score is $-4074.2$. So given a probabilistic model, the probability that tree fits that model is $e^{-4074.2}$, which although low is higher than trees with minor variations that have log likelihood scores in the range $-4075$ to $-4080$.]{
  \epsfig{file=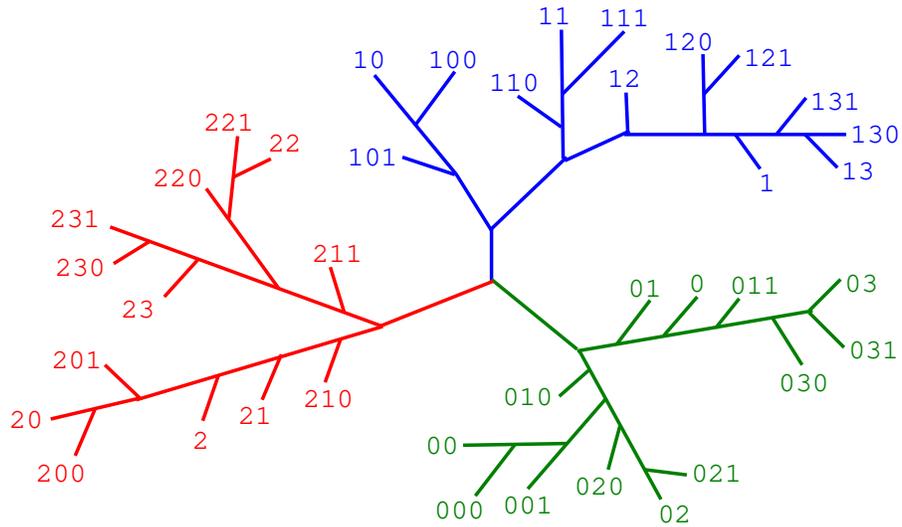,height=7cm}
  \label{trees:maxlikelihood}
  }
  \caption{This shows two trees generated using the parsimony and maximum likelihood methods. A correct tree should have the numbers increasing in size going outwards from the center, with each descendant of $x$ being
  $x\oplus n$, for some $n$.}
  \label{trees}
\end{figure}

After alignment, we then generated the trees shown in Figure~\ref{trees}. We used the PHYLIP software~\cite{Felsenstein96} which implements the two following methods:
\begin{itemize}
\item Parsimony, which generates the tree by evaluating a number of possible trees and finding the one with the overall minimum cost, where the cost is the number of substitutions to explain the observed sequences.
\item Maximum likelihood, which builds a tree with a maximum likelihood of occurring given a model of evolution and the observed sequences. The critical assumptions of the model are that base substitutions (and gaps) follow a Poisson process with a set of specified rates, that each site in the sequence evolves independently, and different lineages evolve independently.
\end{itemize}

Both techniques worked well, however we felt that the parsimony technique worked better than the maximum likelihood technique. A number of open questions have been raised about the efficacy of gap penalty scoring~\cite{FitchandSmith,EmpiricalGaps}. In order to see if the incorrect portions of the tree were due to incorrect scoring of gaps, we repeated the steps outlined above but with no gap-producing insertions and deletions and with a corresponding increase in the number of substitutions (to make the distances comparative). We found no significant difference in the trees produced, suggesting the problems lie elsewhere.
\subsection{Multifractal classification}\label{multi}
Here we use the fractal method as detailed by Yu {\it et al.}~\cite{Multifractal1,Multifractal2}, which considers the
R\'{e}nyi dimension $D_{q}$ for $q \in {\mathbf R}$, 
given by
\begin{equation}
D_{q}=
\begin{cases}
\displaystyle \lim_{\epsilon \to 0}\frac{\ln Z_{\epsilon}(q)}{(q-1)\ln \epsilon}, & q \neq 1,\\
\\
\displaystyle \lim_{\epsilon \to 0}\frac{Z_{\epsilon}(q)}{\ln \epsilon}, & q = 1
\end{cases}
\label{Renyi}
\end{equation}
where 
\begin{equation}
Z_{\epsilon}(q)=
\begin{cases}
\displaystyle \sum_{\mu(B) \neq 0}[\mu(B)]^{q}, & q \neq 1,\\
\\
\displaystyle \sum_{\mu(B) \neq 0}\mu(B)\ln \mu(B), & {\rm otherwise}.
\end{cases}
\label{Z}
\end{equation}
The $\mu(B)$ are simply the sample probabilities (in $[0,1]$) of finding each of the possible substrings
of codons of length $K$, $\epsilon$ is $\epsilon=4^{-K}$. We found a value of $K=8$ works best.
The space $(D_{-1},D_{1},D_{-2})$ used in conjunction with the neighbor joining algorithm~\cite{SaitouNei} is then used to generate phylogenetic trees.

Since the multifractal technique deals with large scale statistics, it is unreliable in dealing with the short sequences we analysed in the previous subsection. The question then arises, is the technique of any use as a whole-genome classification technique? To analyze this question, we model larger changes to bacterial
genomes, in particular both gene transfer~\cite{genetransfer} and shuffling of operons to different positions on the genome~\cite{GeneOrder}. We have focussed on the {\it Escherichia coli} bacteria, due to the considerable amount of detailed information available on their genetic makeup~\cite{EColi} and also the mechanisms by which genetic material can be inserted into their genomes~\cite{ProkaryoticGenetics}. The {\it E.~coli} bacteria we used were K12 (Genbank accession number
NC\_000913), O157:H7 (NC\_002695), O157:H7 EDL933 (NC\_002655), and CFT073 (NC\_004431). In the rest of this paper
we simply use the numerical part of the accession numbers to refer to the different subspecies.
The {\it E.~coli} bacteria have a large, variable portion of foreign genes~\cite{genetransfer} so they provided an excellent test for a technique which is good at picking up differences in gene usage. 

The particular algorithms we used for moving entire genes around are:
\begin{itemize}
\item Gene shuffling. Here we identified operons based on existing information about operons in {\it E.~coli} bacteria which details predicted operons in {\it E.~coli}~\cite{RegulonDB}. Since we wanted to see how the multifractal method copes with shifts in general (it does not in fact distinguish between the different types of shifting that occur), it is 
unimportant if there are some errors in finding operons. For future work on whole genome analysis, especially
that considering the ordering of genes~\cite{GeneOrder}, it may be necessary to significantly improve this rate if
our simulator is to accurately represent actual biological processes.
\item Recombination. Here we identified regions where {\it E.~coli} restriction enzymes could cleave the sequence,
and then picked them at random to insert one or more sequences. Both the enzymes used and the additional sequences could be specified by the user (for example, one could insert the adenosine a2 receptor gene used in the previous subsection and insert that using the Eco57I restriction enzyme~\cite{Eco57I}).
\end{itemize}
As before, we also introduced a number of base substitutions, insertions, and deletions into the sequence, this time we used a random number of each of up to thirty.

\begin{figure}[htbp]
  \centering
  \subfigure[The actual tree is, as generated by our software]{
  \epsfig{file=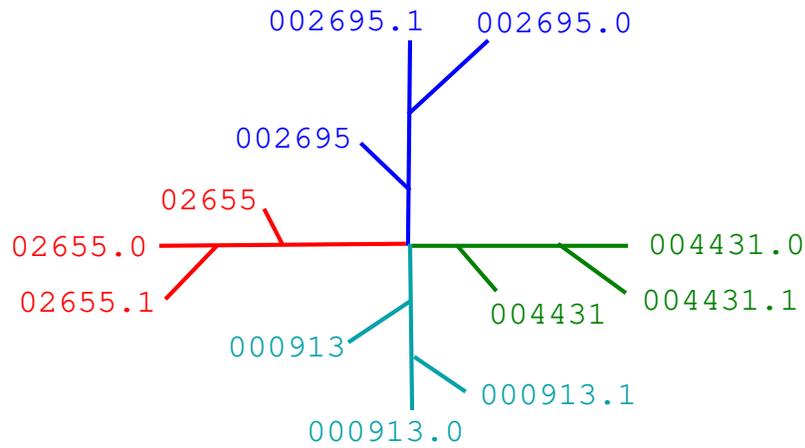,height=6cm}
  \label{mftrees:correct}}
  \subfigure[This is the tree as computed using the multifractal measure and the neighbor joining algorithm. Note that the tree is correct for the 002695 and 000913 families, but has significant problems distinguishing between the 002655 and 004431 families.]{
  \epsfig{file=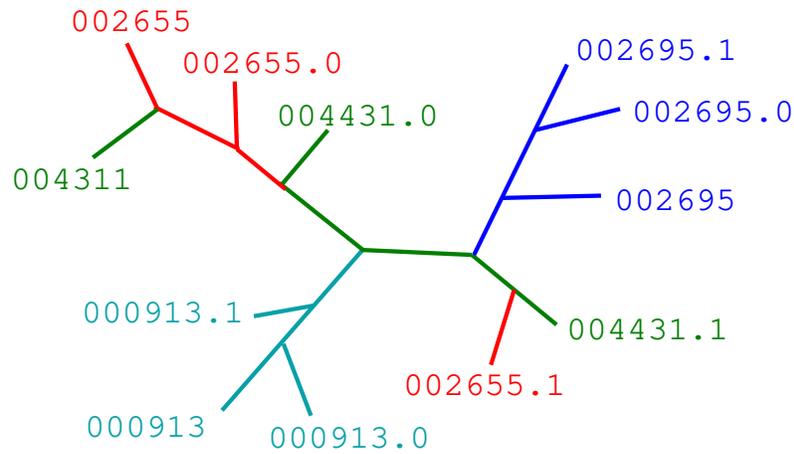,height=6cm}
  \label{mftrees:computed}
  }
  \caption{This shows two trees generated using the parsimony and maximum likelihood methods. A correct tree should have the numbers increasing in size going outwards from the center, with descendants of $x$ labelled $x.n$, $n=0,1$.}
  \label{mftrees}
\end{figure}

We then considered both the parsimony and maximum likelihood methods for comparing a selection of both DNA and amino acid sequences from a set of derived mutants, and used the multifractal with neighbor joining method to compare the whole genome sequences. The results of using the multifractal method this are shown in Figure~\ref{mftrees}.
\subsection{Evalution of multifractal methods}
As can be seen in Figure~\ref{mftrees}, our earlier work~\cite{Berryman1}, and that of Yu {\it et al.}~\cite{Multifractal2}, the multifractal technique has some promise in classifying bacteria and generating phylogenetic trees but has difficulties distinguishing between closely related bacteria. This is due to the fact that it ignores information differences in structure between different genomes and also the features of the genomes such as genes and repeat sequences. To determine an accuracy rate, we simulated large scale genomic changes as in the previous section for both the {\it E.~coli} bacteria previously used as well as for a set of quite different bacteria which consisted of {\it Bacillus cereus}, {\it Shigella flexneri}, {\it Salmonella typhimurium}, {\it Pseudomonas aeruginosa}, {\it Mycoplasma pulmonis}, {\it Lactobacillus plantarum}, and {\it Bradyrhizobium japonicum}. In addition to simply benchmarking against trees produced by our software, we also considered well established relationships between bacteria~\cite{protein_trees,WindsOfChange}. For each tree we determined an accuracy rate by considering the fraction of the tree that was correct and the number of basic tree operations (rotates and shifts) needed to correct the tree. The results are given in Table~\ref{forest} and gave an average of $56\%$ correct.
\begin{table}[htb]
\centering\footnotesize
\caption{The fraction of the organisms with a correct relationship to each other in the trees are shown for a number of simulated sets of evolution and a known set of bacteria. The number of tree operations (shifts and rotates) required to turn the output tree into the correct tree are shown. Simulation run {\bf 8} is for the set of actual bacteria with a known relationship, rather than a simulated set of bacteria.}
\begin{tabular}{|c|c|c|}\hline
{\bf Simulation run} & {\bf \% of tree in correct positions} & {\bf Tree op's required to correct}\\\hline
{\bf 1} & 50\% & 4 \\\hline
{\bf 2} & 33\% & 7 \\\hline
{\bf 3} & 58\% & 5 \\\hline
{\bf 4} & 83\% & 2\\\hline
{\bf 5} & 50\% & 7 \\\hline
{\bf 6} & 42\% & 5 \\\hline
{\bf 7} & 75\% & 3 \\\hline
{\bf 8*} & 57\% & 4 \\\hline
\end{tabular}
\label{forest}
\end{table}

Overall the multifractal technique is useful in determining rough relationships between organisms, but has trouble coping with finer details, most likely because it doesn't use enough of the information available about the genomes in question. In the following subsection we detail a related information theory approach.
\subsection{Asymptotically optimal universal compression metrics}
Here we use the same set of files we generated in subsection~\ref{multi} and apply a lossless compression algorithm, with a coding rate that approaches the Shannon rate~\cite{zip1,zip2}:
\begin{equation}
H\left(\mathit{P}\right)=-\displaystyle \sum_{p_{i} \in \mathit{P}} p_{i}\log p_{i}
\label{Shannon}
\end{equation}
The Shannon entropy is in fact a special case of the more general R\'{e}nyi dimension~\ref{Renyi} used in Subsection~\ref{multi}.

\begin{figure}[htbp]
  \centering
  \epsfig{file=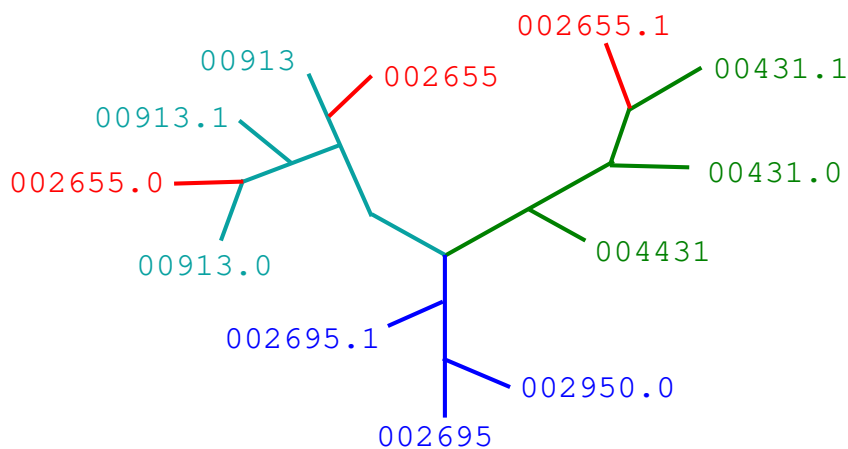,height=6cm}
  \caption{This is the tree produced by using an information theory based metric in conjunction with the neighbor joining algorithm. As with the multifractal tree, it has trouble distinguishing between closely related genomes, for similar reasons. We propose tightening the tree building decision methods using techniques outlined in Gutman~\cite{Gutman}. This will enable us to place accuracy levels on parts of the tree.}
  \label{infotree}
\end{figure}

We use the metric
\begin{equation}
M=H(X|Y)^{2}=\left(H(X)-H(X,Y)\right)^{2},
\label{joint}
\end{equation}
where 
\begin{equation}
H(X,Y)=\displaystyle \lim_{(n,N)\rightarrow \infty} \frac{E\left[L\left(x_{1}^{n}\oplus y_{1}^{N}\right)\right]}{n+N}
\label{oplus}
\end{equation} and
\begin{equation}
H(X)=\displaystyle \lim_{n\rightarrow \infty} \frac{E\left[L\left(x_{1}^{n}\right)\right]}{n}.
\label{hx}
\end{equation}
In Equations~\ref{oplus} and~\ref{hx}, $x_{1}^{n}$ and $y_{1}^{N}$ are the sequence strings of the two genomes of lengths
$n$ and $N$ bits (two bits per base), and $E\left[L\left(\cdot \right)\right]$ is the compressed length operator (again, in units of bits). The tree for the same mutations as in Subsection~\ref{multi} is shown in Figure~\ref{infotree}. As with the multifractal measure, the information theory measure has difficulties determining differences between closely related bacteria.
\section{Conclusions}
In general we found that the multifractal and compression metrics we have explored are useful in distinguishing between different organisms, they have problems distinguishing between those which are closely related, or those which are too small for statistical properties to be estimated with any accuracy. We found the popular maximum likelihood and parsimony techniques to handle small changes in short sequences quite well, with the parsimony tree building method handling gaps better than maximum likelihood. We used an entirely automated approach to the problem of multiple alignment, using the CLUSTALW software package. A skilled operator would be able to edit the CLUSTALW results to produce better alignments which would help the maximum likelihood and parsimony algorithms further. 

Although the multifractal and compression metrics were found to be useful, the fact that they ignore a considerable amount of whole-genome information available such as gene reversals and gene ordering between different species means they will always have certain limitations in accuracy. Work can be done to specify the accuracy of the metrics produced by these methods. In particular, the compression method can identify organisms that it is unable to classify reliably and this could be indicated when drawing trees.
the way ahead in whole genome comparisons and the resultant phylogenies lies with combining multiple techniques including, but not limited to, comparison of sets of aligned protein sequences, comparison of sets of {\it unaligned} protein sequences~\cite{unalignedproteins}, and information on gene order changes~\cite{GeneOrder,GeneOrderB}. Some of these techniques can be time consuming, however, requiring significant amounts of human input. Perhaps the best approach would be to combine whole genome techniques with those of orthologous gene comparisons and even physiological data in determining highly accurate phylogenies. 
\acknowledgments
We gratefully acknowledge funding from The University of Adelaide.
\bibliography{phd}   
\bibliographystyle{spiebib}   
\end{document}